# Record high bandwidth integrated graphene photodetectors for communication beyond 180 Gb/s

Daniel Schall[1], Emiliano Pallecchi[2], Guillaume Ducournau[2], Vanessa Avramovic[2], Martin Otto[1], Daniel Neumaier[1]

[1]*Advanced Microelectronic Center Aachen, AMO GmbH, Otto-Blumenthal-Str. 25, 52074 Aachen, Germany*
[2]*Institut d'Electronique de Microélectronique et de Nanotechnologie (IEMN), UMR CNRS 8520, Université de Lille 1, 59652 Villeneuve d'Ascq CEDEX, France.*
schall@amo.de

**Abstract:** We report on the fastest silicon waveguide integrated photodetectors with a bandwidth larger than 128 GHz for ultrafast optical communication. The photodetectors are based on CVD graphene that is compatible to wafer scale production methods.

## 1. Introduction

In the recent past graphene was established as photonic material that allows realizing a variety of novel optic and electro optic functionalities and device architectures [1]. The key features of graphene that distinguish it from other semiconductor materials are the integrability on almost any substrate and the extraordinary linear band structure that leads to a flat absorption characteristic from the UV to the far infrared and moreover enables ultrafast carrier dynamics. Upon optical excitation, charge carriers thermalize within 50 fs which leads to an intrinsic bandwidth in the THz range [2]. The fastest measured extrinsic bandwidths so far for silicon waveguide integrated graphene devices are 40 GHz [3,4], 65 GHz [5] and 76 GHz [6]. These bandwidths already reveal the potential of graphene for ultrafast optical communication, however, they remain behind the bandwidth of 120 GHz of the best silicon waveguide integrated germanium photodetectors [7]. Here we report on graphene photodetectors that have a bandwidth larger than 128 GHz suitable beyond 180 Gb/s of optical data transmission rate in the most simple on-off-keying scheme.

## 2. Experiment

The silicon photonic structures designed for 1550 nm wavelength were fabricated on 150 mm silicon-on-insulator wafers. Waveguide integrated photodetectors were subsequently fabricated form large scale chemical vapor deposited graphene. A microscopic top view of a device is shown in figure 1a. The graphene layer is located in the evanescent field of the waveguide and thereby absorbing the light. Figure 1b shows the characterization of the photonic layer without graphene photodetectors. The loss of the grating couplers is 3.5 dB and the total optical loss of the grating couplers and the waveguide to the photodetector is 3.7 dB on average, see histogram in the inset of figure 1b for a typical distribution one die. The 0.3 dB spread of the histogram is mainly given by the fiber-to-chip alignment tolerances, rather than by the variation of the photonic structures.

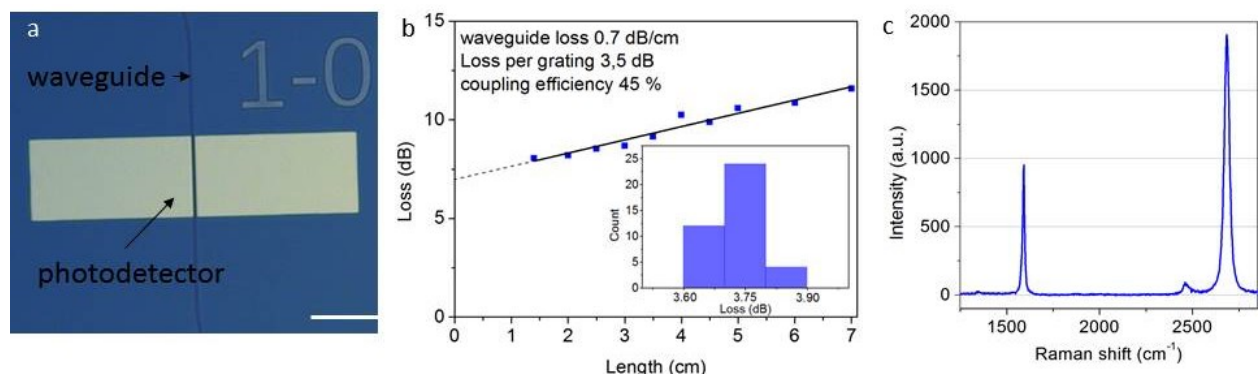

Figure 1 a) Microscopic picture of a graphene photodetector on a silicon waveguide. The scale bar is 50 µm. b) Optical TLM measurements to extract the losses of the grating couplers (3.5 dB) and the waveguides (0.7 dB/cm). The inset shows a histogram with optical losses for the grating coupler and the waveguide to the photodetector measured on one die. c) Typical Raman spectrum of graphene from the same batch.

Raman measurements were performed on reference samples from the same batch to determine the quality of the monolayer graphene. Figure 1c shows a typical Raman spectrum with the G peak at 1591 cm$^{-1}$, the 2D peak FWHM of 29.2 cm$^{-1}$ and a negligible D peak. The charge carrier mobility was approximately 2000 cm²/Vs measured in a separate experiment on graphene from the same batch. The responsivity of the graphene photodetectors was measured using a modulated light source (60 kHz) at 1550 nm and a lock-in amplifier that is connected directly to the photodetectors. For the RF measurements a heterodyne setup described in [5] was used. All measurements were performed at room temperature under ambient conditions.

## 3. Results and discussion

The designed resistance of the graphene photodetectors is matched to the 50 Ohm measurement environment and varied in the as-fabricated devices from 49 to 117 Ohm. Figure 2a shows the photo voltage for a typical device at 0.5V bias as a function of the optical power available at the photodetector after subtracting the photonic losses. The photo voltage follows a square root like shape which has been described before by Shiue et al [4]. The signal amplitude is 4.27 mV$_{pp}$ for an input power of 0.28 mW$_{pp}$ at a bias of 0.5 V, the corresponding voltage responsivity is 15 V/W. The maximum current responsivity we reach is 0.18 A/W at 0.5V bias. The average responsivity at a bias of 0.25 V for 10 measured devices is 32 mA/W, see inset in figure 2a. The DC response is the highest so far reported for graphene photodetectors fabricated using wafer scale material. In comparison to our earlier results on wafer scale fabricated devices the responsivity was increased by two orders of magnitude [6].

In the next step we determined the RF response of the photodetectors at 1 GHz optical input signal with a power of 19.6 dBm. The maximum RF output power of -25 dBm was observed at a bias voltage of 2 V for a device with a resistance of 116 Ω. This RF output power is the highest reported for graphene devices [3-6]. Up to now, the highest output power was -32 dBm which was however obtained by non-scalable graphene-flake based photodetectors [5].

The frequency dependence of the output power was measured by tuning the beating frequency of the heterodyne system between 1 GHz and 128 GHz. A typical characteristic is shown in the plotted frequency range from 1 to 128 GHz in figure 2b for a bias voltage of 1 V. The measured signal was calibrated by subtracting the frequency characteristic of the cable, bias-tee and power meter. The scattering parameters of these RF components and the calibration factor of the power meter were measured separately up to 110 GHz. Above 110 GHz, we used the calibration factor measured at 110 GHz to correct the measurement data up to 128 GHz for the minimum losses that can be expected in this 18 GHz frequency range.

For correction of the GS wafer prober we used S$_{21}$ data available up to 67 GHz. This part of the characteristic is shown in blue in figure 2b. Above 67 GHz, calibration substrates are not available for GS probe tips. The loss of the probe tip is approximately 1.6 dB at 67 GHz, which has been subtracted as lower bound of the tip loss between 68 and 128 GHz (grey in figure 2b).The dips and peaks in the grey part of the frequency response are due to resonances in the power meter and the RF connectors and slight frequency uncertainties related to the heterodyne signal generation. We note that the average output power does not drop below -3 dB up to the end of our measurement range of 128 GHz.

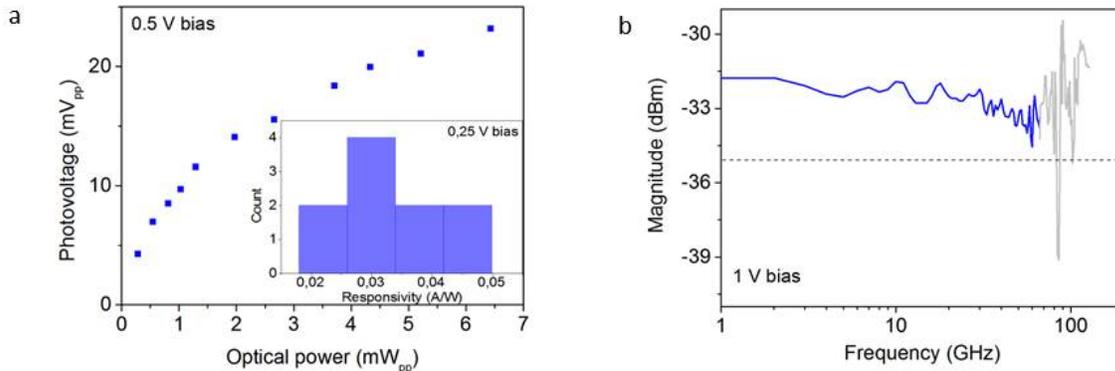

Figure 2 a) Power dependence of the photo voltage measured at a bias of 0.5 V. The optical power is corrected for the losses due to the grating coupler and waveguide. b) Frequency response of a typical photodetector measured at 1 V bias. The bandwidth is larger than the measurement range of 128 GHz. The frequency axis is plotted up to 200 GHz.

The bandwidth of the photodetectors is larger than 128 GHz, the highest value so far for graphene photodetectors. Moreover, the fastest germanium photodetector that is integrated on a silicon waveguide has a 3 dB bandwidth of 120 GHz [7]. The characteristic shown in figure 2b has not yet started to roll off at 128 GHz, indicating that the bandwidth is significantly larger than 120 GHz. Note, that the frequency axis is plotted up to 200 GHz. Hence we present here the fastest integrated photodetector available today for integrated silicon photonic RF systems. The bandwidth larger than 128 GHz is sufficient for data transmission faster than 180 Gb/s in the most simple on-off-keying scheme.

## 4. Summary

We report on the fastest silicon waveguide integrated photodetectors with a bandwidth larger than 128 GHz for ultrafast optical communication. The photodetectors are based on CVD graphene that is compatible to wafer scale production methods.

## 5. Acknowledgement

This work was funded by the European Commission under the contract no. 696656 ("Graphene Flagship"). We thank R. Negra for providing us with RF equipment.

[1] F. H. L. Koppens, T. Mueller, Ph. Avouris, A. C. Ferrari, M. S. Vitiello and M. Polini Photodetectors based on graphene, other two-dimensional materials and hybrid systems Nat. Nanotech. 9, 780 – 793 (2014)

[2] K. J. Tielrooij, L. Piatkowski, M. Massicotte, A. Woessner, Q. Ma, Y. Lee, K. S. Myhro, C. N. Lau, P. Jarillo-Herrero, N. F. van Hulst, F. H. L. Koppens Generation of photovoltage in graphene on a femtosecond timescale through efficient carrier heating Nat. Nanotech. 10, 437-443 (2015)

[3] D. Schall, D. Neumaier, M. Mohsin, B. Chmielak, J. Bolten, C. Porschatis, A. Prinzen, C. Matheisen, W. Kuebart, B. Junginger, W. Templ, A. L. Giesecke and H. Kurz "50 GBit/s Photodetectors Based on Wafer-Scale Graphene for Integrated Silicon Photonic Communication Systems" ACS Photonics 1, 781-784 (2014)

[4] R. J. Shiue, Y. Gao, Y. Wang, C. Peng, A. D. Robertson, D. K. Efetov, S. Assefa, F. H. L. Koppens, J. Hone and D. Englund "High-Responsivity Graphene Boron Nitride Photodetector and Autocorrelator in a Silicon Photonic Integrated Circuit" Nano Lett. 15, 7288-7293(2015)

[5] S. Schuler, D. Schall, D. Neumaier, L. Dobusch, O. Bethge, B. Schwarz, M. Krall, T. Mueller "Controlled Generation of a p–n Junction in a Waveguide Integrated Graphene Photodetector" Nano Lett. 16, 7107–7112 (2016)

[6] D. Schall, C. Porschatis, M. Otto and D. Neumaier "Graphene photodetectors with a bandwidth >76 GHz fabricated in a 6" wafer process line" J. Phys. D: Appl. Phys. 50 124004 (2017)

[7] L. Vivien, A. Polzer, D. Marris-Morini, J. Osmond, J. M. Hartmann, P. Crozat, E. Cassan, C. Kopp, H. Zimmermann, J. M. Fédéli, "Zero-bias 40 Gbit/s germanium waveguide photodetector on silicon" Optics Expr. 20, 1096 (2012)